\documentclass[manuscript]{aastex}
\usepackage[footnotesize]{caption}

\begin{document}

\title{Single or Double Degenerate Progenitors? Searching for Shock Emission in the SDSS-II Type Ia Supernovae}
\author{Brian T. Hayden\altaffilmark{1},
Peter M. Garnavich\altaffilmark{1},
Daniel Kasen\altaffilmark{2,3},
Benjamin Dilday\altaffilmark{4,5,6},
Joshua A. Frieman\altaffilmark{7,8,9},
Saurabh W. Jha\altaffilmark{4},
Hubert Lampeitl\altaffilmark{10},
Robert C. Nichol\altaffilmark{10},
Masao Sako\altaffilmark{11},
Donald P. Schneider\altaffilmark{12},
Mathew Smith\altaffilmark{13},
Jesper Sollerman\altaffilmark{14},
J. Craig Wheeler\altaffilmark{15},
Richard Kessler\altaffilmark{7,8,16}
}
\altaffiltext{1}{University of Notre Dame, 225 Nieuwland Science Hall, Notre Dame, IN 46556, USA}
\altaffiltext{2}{Department of Physics, University of California, Berkeley}
\altaffiltext{3}{Lawrence Berkeley National Laboratory, Berkeley, Ca, 94720, USA}
\altaffiltext{4}{Department of Physics and Astronomy, Rutgers University, 136 Frelinghuysen Road, Piscataway, NJ 08854, USA}
\altaffiltext{5}{Las Cumbres Observatory Global Telescope Network, 6740 Cortona Dr. Suite 102, Santa Barbara, CA 93117, USA}
\altaffiltext{6}{Department of Physics, University of California, Santa Barbara, Broida Hall, Mail Code 9530, Santa Barbara, CA 93106-9530, USA}
\altaffiltext{7}{Department of Astronomy and Astrophysics, The University of Chicago, 5640 South Ellis Avenue, Chicago, IL 60637, USA}
\altaffiltext{8}{Kavli Institute for Cosmological Physics, The University of Chicago, 5640 South Ellis Avenue, CHicago, IL 60637, USA}
\altaffiltext{9}{Center for Particle Astrophysics, Fermi National Accelerator Laboratory, P.O. Box 500, Batavia, IL 60510, USA}
\altaffiltext{10}{Institute of Cosmology and Gravitation, University of Portsmouth, Portsmouth PO1 3FX, UK} 
\altaffiltext{11}{Department of Physics and Astronomy, University of Pennsylvania, Philadelphia, PA 19104, USA }
\altaffiltext{12}{Department of Astronomy and Astrophysics, 525 Davey Laboratory, Pennsylvania State University, University Park, PA 16802, USA} 
\altaffiltext{13}{Astrophysics, Cosmology and Gravity Centre (ACGC), Department of Mathematics and Applied Mathematics, University of Cape Town, Rondebosch, South Africa. 7700}
\altaffiltext{14}{The Oskar Klein Centre, Department of Astronomy, Albanova, Stockholm University, SE-106 91 Stockholm, Sweden}
\altaffiltext{15}{Department of Astronomy, University of Texas, Austin, TX 78712, USA}
\altaffiltext{16}{This author was inadvertently not included in the journal version of this paper}

\begin{abstract}
From the set of nearly 500 spectroscopically 
confirmed type~Ia supernovae and around 10,000 unconfirmed candidates from SDSS-II, we select a subset of 108 confirmed SNe Ia with well-observed early-time light curves to search for signatures from shock interaction of the supernova with a companion star. No evidence 
for shock emission is seen; however, the cadence and photometric noise could hide a weak shock signal. 
We simulate shocked light curves using SN Ia templates and a simple, Gaussian shock model to emulate the noise properties of the SDSS-II sample and estimate the detectability of the shock interaction signal as a function of shock amplitude, shock width, and shock fraction. 
We find no direct evidence for shock interaction in the rest-frame $B$-band, but place an upper limit on the shock 
amplitude at 9\% of supernova peak flux ($M_B> -16.6$ mag). If the single degenerate channel dominates type~Ia 
progenitors, this result constrains the companion stars to be less than about 6 $M_{\odot}$ on the main sequence, and strongly disfavors red giant companions.  
\end{abstract}

\keywords{supernovae: general} 

\section{Introduction}
Although Type Ia supernovae (SNe Ia) have been used as standard candles for around 15 years
\citep{phil93,ham95,riess95}, the nature of the progenitor for these objects remains elusive.
It is widely accepted that they are thermonuclear explosions of white dwarfs (WD) that are
nearing the Chandrasekhar limit. The nature of the mass gain is generally categorized
into two scenarios: the single degenerate (SD) scenario \citep{whe73, nom82},
where the WD accumulates mass from a binary companion, and the double degenerate (DD)
scenario \citep{web84, iben84} in which two white dwarfs coalesce. Although the single-degenerate
channel is widely recognized as the most plausible, the lack of hydrogen and helium in SN Ia spectra is a mystery \citep{liv00,mat05,leo07,kotak08}. 

A WD in the SD channel must burn the accreted material in order to avoid
classical nova outbursts, therefore these systems are expected to be X-ray sources \citep{van92}.  A
recent X-ray study \citep{gil10} found that the observed X-ray flux from 6 nearby elliptical
galaxies is an order of magnitude lower than the value expected if all SNe Ia occurred in the
SD channel, under the simplifying assumption of a constant duty cycle. They conclude that the SD channel can account for, at most, 5\% of Type Ia
explosions in these environments. Based on SN Ia rates (e.g. \citealt{dil10}), \citet{dis10} calculated that
several hundred supersoft sources should be detectable in galaxies similar to M31 and the Milky Way; 
the actual number of supersoft sources is about two orders of magnitude below this. A similar analysis
was extended to the DD scenario in \citet{dis10b}, and again, too few supersoft sources are detected
given the well-defined rate of SNe Ia. \citet{dis10b} suggests that not all nuclear-burning WDs
(white dwarfs that are burning and retaining material transfered by a companion) are visible as
supersoft sources. An example could be the recurrent novae that are nuclear-burning SD systems that do not necessarily spend an appreciable fraction of their accretion phase as supersoft sources \citep{sch10}. \citet{pan06} studied $\sim 30$ SNe Ia at radio wavelengths and were unable to detect a significant signal, arguing against the SD model with massive companions but not ruling out smaller companions or the DD scenario.  Although the energetics of some bright type~Ia events such as SN 2006gz \citep{hic07}, 
SN 2003fg \citep{how06}, SN 2007if \citep{sca10}, and SN 2009dc \citep{sil10,yam09} may require the DD channel because of inferred ejecta masses larger than the Chandrasekhar limit,
the progenitor systems of normal SNe~Ia remain uncertain. 

\citet{kas10} predicted that the interaction of the exploding WD with its companion star (in the SD channel)
should be analogous to shock breakout in core collapse SNe.  Kasen expects the shock signature
to appear less than a day after explosion as X-ray emission as the companion star carves a hole in the optically thick early-time supernova ejecta. The shock will also produce excess emission at
ultraviolet and blue wavelengths. The size of 
the companion and the separation of the stars controls the amplitude of the shock. For compact companions
filling their Roche lobe, the shock emission in the optical is a
small fraction of the peak supernova flux, while a red giant companion would result in shock
emission comparable to the supernova at peak. Due to the solid angle of the
interaction,
approximately 10\% of SNe~Ia that occur in the SD channel should show signatures of this interaction. Further discussion of the nature of the interaction as well as the potential observational signatures can be found in \citet{mar00} and \citet{pak08}.

This article analyzes a large number of type~Ia SN light curves from the SDSS-II Supernova Survey \citep{frie08, sdsscosmo}
for evidence of the interaction. To date, no evidence for shock emission at early
times has been reported, despite the discovery of hundreds of SNe Ia. Until recently, however,
early observations of type~Ia events have been sporadic and the orientation effects are expected
to significantly reduce the number of SD events that would show visible shock emission.
It is also possible that the SD channel is not the sole or even dominant channel of SN Ia production,
which would greatly decrease the potential number of observable shocks. The SDSS-II Supernova Survey \citep{frie08}
provides a large-area,  deep search with a rapid cadence. The nature of the search (viewing the same patch of sky every couple of days) resulted in hundreds of observations within a few days of explosion, and permits a well-defined statistical test
for the presence of shocks.

In section 2, we briefly describe the SDSS-II data and SNANA simulated light curves,
and explain how we simulate the shock interaction in the context of the \citet{kas10} model.
Section 3 explains our chosen method of statistically testing for the presence of shock emission.
Section 4 is a brief discussion of our results, followed by our conclusions in section 5. 

\section{Light Curves: Real and Simulated}
\subsection{Data}
This study incorporates 498 spectroscopically confirmed SN~Ia light curves along with 10,400 unconfirmed transient objects from the SDSS-II Supernova Survey \citep{frie08}. The SDSS-II Supernova Survey conducted three observing campaigns between September and December in each of 2005, 2006, and 2007 using the 2.5 m telescope at Apache Point Observatory \citep{gunn06} and the SDSS camera \citep{gunn98}. The search was designed to scan 300 deg$^2$ of sky as often as every-second night, in order to find several hundred SNe Ia at intermediate redshifts (z $\approx$ 0.2); this redshift range had been sparsely observed and populating it with a few hundred SNe Ia allows for the testing of many cosmological models. Each candidate supernova was inspected visually to rule out image artifacts and asteroids. The objects were targeted for spectrographic observation based on their probability of being SNe Ia \citep{sako08}.  The SDSS-II SN Survey is aided by the extensive database of reference images, object catalogs, and photometric calibration compiled by the SDSS \citep{york00}.

In this study, we analyzed $\sim$10,400 of the transient objects discovered by the SDSS-II but not confirmed as SNe Ia. These objects were fit twice by the light curve photometric typing algorithm described by \citet{sako08}. In the first fitting process, all data was allowed in the fit (resulting probability referred to here as $P_{Ia,i}$), while in the second fitting process, data earlier than 
$-10$ days was ignored ($P_{Ia,f})$. It is possible that by classifying these objects with a light curve fitting process, objects containing early shock emission may be rejected as SNe Ia since they do not match a normal SN Ia template. The \citet{sako08} algorithm provides a probability that the object is a SN Ia; we selected 82 objects ($\sim$ 0.7\%) with $P_{Ia,f} - P_{Ia,i} \geq 0.95$ and analyzed each light curve by visual inspection. This sample contains only objects that were $\leq 5\%$ probability before and $\geq 95\%$ probability after cutting the early data. None of these objects is a shock candidate based on visual inspection. We conclude that the typing process did not reject any SNe Ia with shocks. 

\subsection{Shock Simulation}
This study also utilizes approximately 3000 simulated light curves created using SNANA \citep{kess09}. SNANA provides simulated light curves with the same cadence, signal-to-noise, and light curve width distribution as the SDSS-II data (as well as many other surveys). For our implementation, we use the MLCS2k2 \citep{MLCS} template in SNANA with a 16.5 day rise time extrapolation \citep{hay10}.  Extrapolation is necessary because MLCS2k2 is trained only on data later than 10 days before peak brightness. Based on recent studies on the shape of the early light curve \citep{hay10, con06}, we use a $t^2$ extrapolation consistent with the \citet{arnett82} prediction. Note that this 16.5 day extrapolation does not represent the average rise time of a SN Ia, but the best relative shape of the transition between the extrapolation and the MLCS template. The shocks are expected to be blue, so we only analyze the rest-frame $B$-band light curves for this study. 

For light curve fitting, we use the 2-stretch method developed by \citet{hay10}, with two differences. First,
we begin the fitting process at 10 days before maximum to avoid including shock emission in the fit; \citet{hay10} used all available pre-maximum points in the fitting process. As a consequence, the template extrapolation to explosion has no effect on our fits in this analysis, however, we use the 16.5 day extrapolation in Figure 1.
The second difference is that after fitting all available light curves, we cut out those light curves with rise or
fall time errors greater than 4 days (up from 2 days, in \citealt{hay10}). This is more lenient than the analysis in \citet{hay10} because
restricting the fit to greater than $-10$ days increases the uncertainty in the rise-time and lowers
the number of events passing the cut. As we are more interested in finding shocks than accurately measuring
the rise time, we have eased the uncertainty requirement. 108 SDSS-II SNe Ia pass our error cuts, with a redshift range of $0.037 - 0.278$ and a median redshift of 0.157. 695 of the SNANA simulated light curves pass our cuts.

The light curves predicted by \citet{kas10} are the result of shock emission due to the supernova ejecta colliding with the companion star (see Figure 1 of \citealt{kas10}). At early times, the supernova photosphere
is optically thick; this conical hole provides a window for radiation to quickly escape in the form of X-rays.
The interaction is visible in a solid angle of  $\Omega/4\pi \approx 1/10$, which means that about $10\%$ of SN~Ia
produced through the SD channel would result in an observable light curve signature. That is, even if all
SN~Ia came from SD progenitors, at best, only $\sim 10$\% would have an associated shock emission
visible to an observer on Earth.

For this analysis, we chose to simulate the shock emission. Analyzing individual light curves for shock emission is difficult for many reasons. There is noise in the early data and the time of explosion is not measured particularly well when the fit is restricted to points beyond -10 days. Using composite light curves containing hundreds of points helps to reduce the impact of the fitting error by allowing for robust statistical testing. With the SDSS-II Type Ia light curves, analyzing simulated shocks was the most robust way to set limits on shock emission in the data. Analyzing individual light curves without using simulated shocks would require small fit errors and an even more rapid cadence, in order to get several observations including shock emission (allowing for robust statistical testing) and not just one. It is also necessary for us to produce simulated shocks in order to populate the parameter space, as \citet{kas10} provides three theoretical curves. To incorporate shocks into simulated light curves, we first fit the SNANA generated light curves (in $B$ band) with the 2-stretch fitter \citep{hay10}. As with the real data, the fit is restricted to the period from 10 days before peak to 25 days after.
For both the data and the simulations, supernovae with rise time and fall time errors greater than 4 days
are excluded. We then 2-stretch correct the simulated curves (divide the rise portion by the fitted rise stretch and the fall portion by the fitted fall stretch) so that the light curves match the MLCS2k2 template that
was used in the fitting process. At this point in the process, all of the simulated light curves are in the supernova rest-frame and have been 2-stretch corrected to the template. For a fraction of these light curves, we then add the fading
half of a Gaussian function to the simulated data at the time of explosion, creating a simulated shock. We use this Gaussian because it allows greater flexibility in the magnitude and width of the shock. \citet{kas10} provides three shocked light curves, so our Gaussian method allows population of the parameter space for shocks on a much finer scale.
The beginning of the shock (the peak of the Gaussian) is allowed to fluctuate with a 0.5-day standard deviation
to allow for the uncertainty in the time of explosion. There are three variables that we consider in this study:
the width of the Gaussian light curve (ranging from 0.5 to 5 days, referred to as ``shock width'' or $\sigma$),
the amplitude of the shock (as a percent of maximum flux, referred to as ``shock amplitude'' or $S$),
and the fraction of simulated SNe that are shocked (5\%, 10\%, or 20\%, referred to as ``shock fraction''). 
The percentage of visible shocks is the product of the fraction of SNe~Ia that come from SD progenitors and the 
fractional solid angle of shock emission.

\begin{figure}[h!]
\begin{tabular}{cc}
\includegraphics*[scale=0.5]{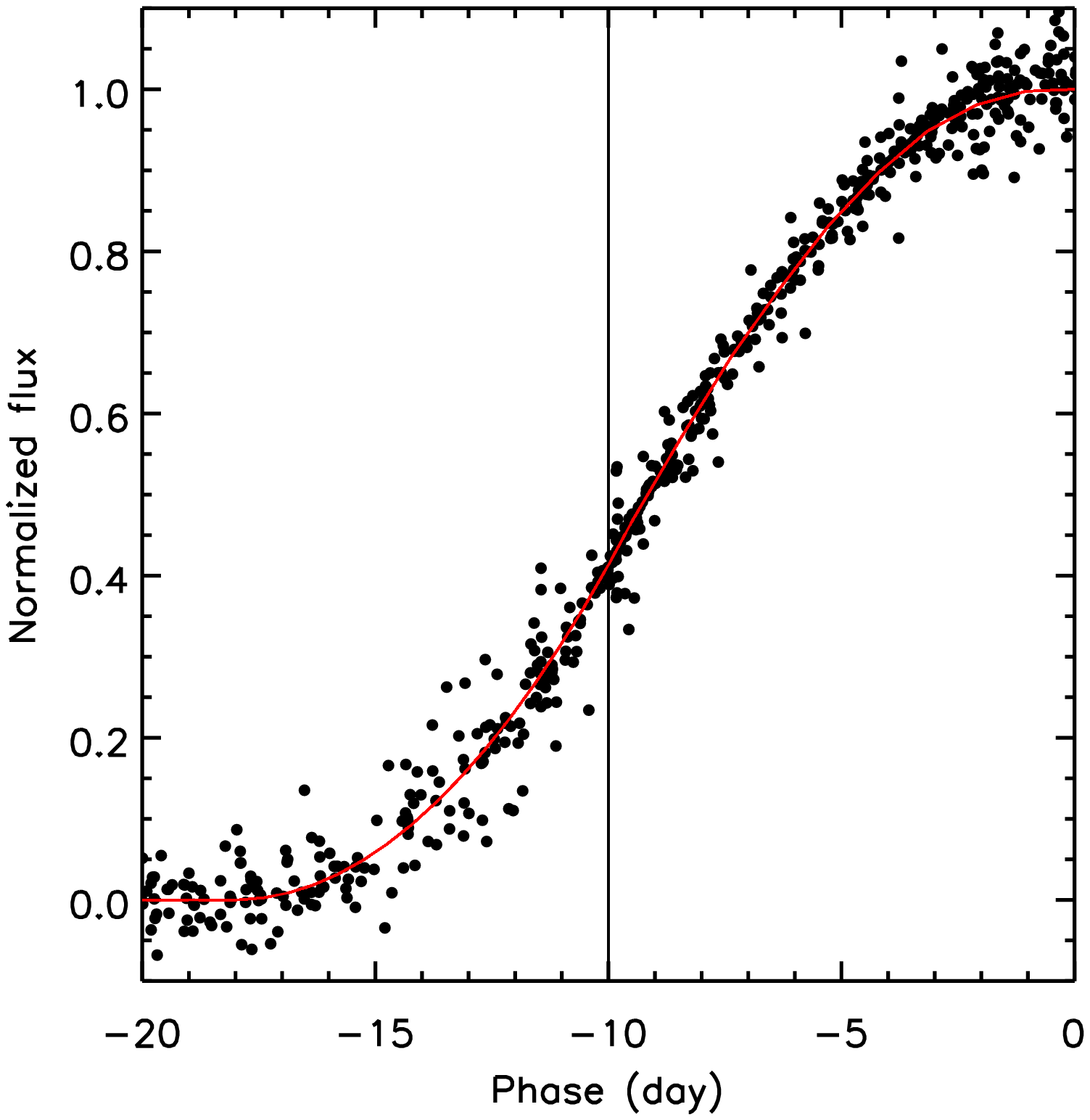} & \includegraphics*[scale=0.5]{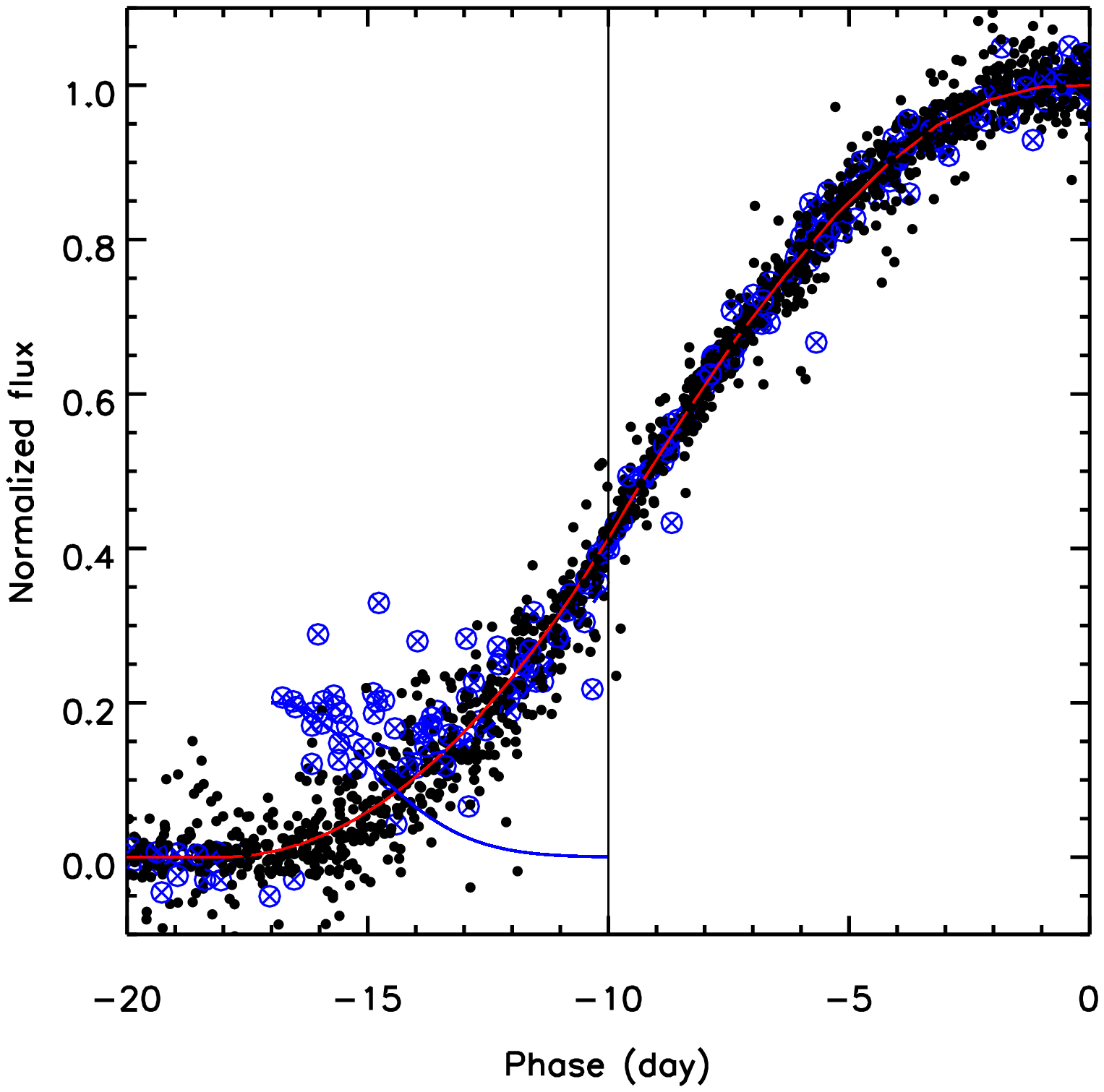}
\end{tabular}
\caption{{\it Left:} 108 type~Ia SNe from the SDSS-II Supernova Survey normalized in peak flux ($B$ band) and rise stretch
to the template. The vertical line shows the cutoff phase for
the fitting process; no points earlier than $-10$ days were included in the fits.  {\it Right:} simulated
light curves from SNANA, matching the cadence and signal-to-noise of SDSS-II. The $\sim 350$ simulated curves shown
here were fit with the 2-stretch fitter, then 2-stretch corrected to the template. The unshocked light curves are shown as black dots, with the template
drawn in red. The blue circles indicate the 10\% of supernovae which had shocks added. The simulated shock
emission is shown as a blue line and has a width of 2~days and an amplitude of 20\% of the peak SN flux. This represents an example of one simulated set of shocks with a 10\% shock fraction, 2 day shock width, and 0.2 shock amplitude. Errors are not included; see \citet{hay10} for similar SDSS-II light curves with error bars included. The shocked light curves in blue add a noticeable dispersion to the early light curve. The Brown-Forsythe test indicates a statistically significant difference in the variances between this set and an unshocked set of simulated light curves. \label{shocks}}
\end{figure}

Figure \ref{shocks} shows the SDSS-II light curves used in this analysis, as well as an example of simulated
shocks in the SNANA light curves. In the left panel, the SDSS-II light curves (in the $B$ band) have been 2-stretch corrected
(using only data between $-10$ and 25 days in the fit) to the template. At first glance, the data indicate that shocks are not clearly present
in the SDSS-II data when compared to the SNANA light curves simulated with the same cadence and noise properties;
there are no obvious shocks in the 108 real light curves passing the selection cuts. The right panel shows
approximately 350 simulated light curves randomly selected from the full sample of 695, and treated in the same manner as the real data. The blue circles represent
the 10\% of SNe that were given a simulated shock (i.e. a shock fraction of 10\%). The template light curve
is the red line, and the added shock is the blue line. The blue points, the points that have been given a simulated shock, are simply the red line plus the blue line, with the noise from the data. The blue points display a clear
increase in variance due to the simulated shock, which in this particular case adds 20\% of the SN peak flux at the time of explosion 
and then fades as a Gaussian with a width of 2~days. To quantify the detection of shocks, we test for a difference in variance 
between ``shocked" and ``unshocked" samples using the Brown-Forsythe test \citep{bf74}, described in the next section. 
A statistical test of the example distribution shown in the right pane of Figure \ref{shocks} indicates that it is significantly 
different from an un-shocked sample (probability of the same variance is less than 5\%); we would consider this a
robust ``detection" of shocks.

\begin{figure}[h!]
\begin{tabular}{cc}
\includegraphics*[scale=0.4]{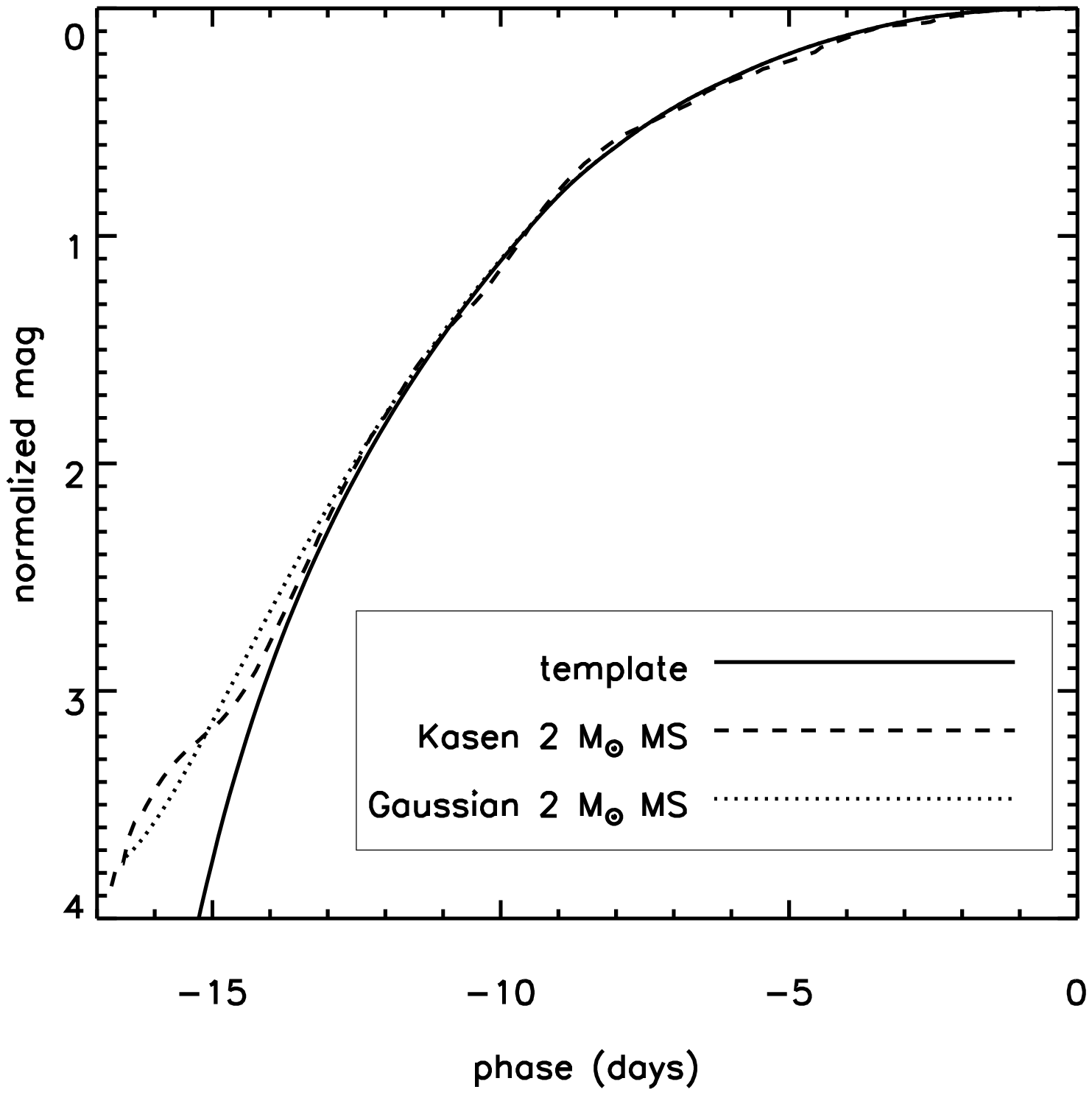}&\includegraphics*[scale=0.4]{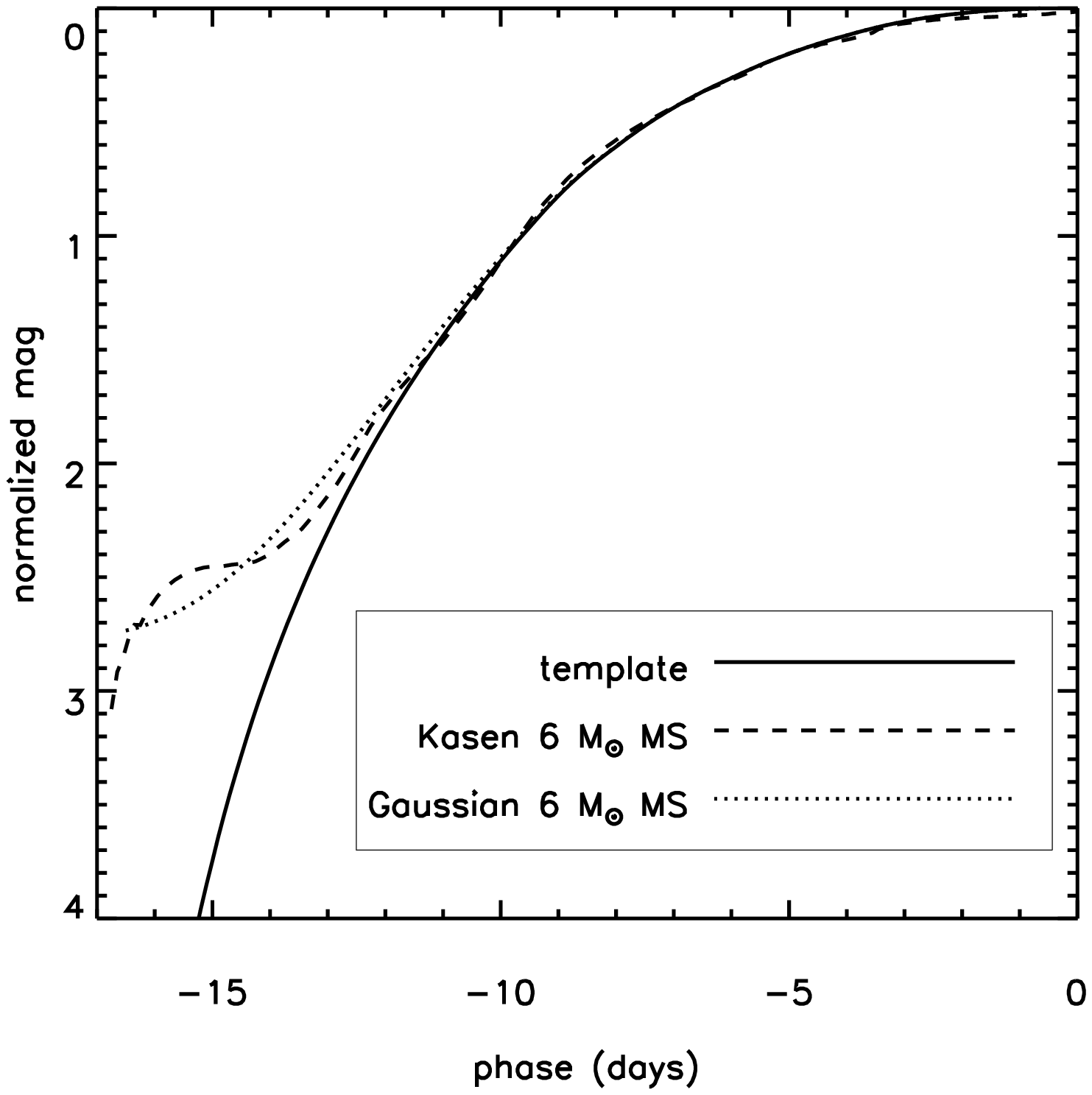}\\
\includegraphics*[scale=0.4]{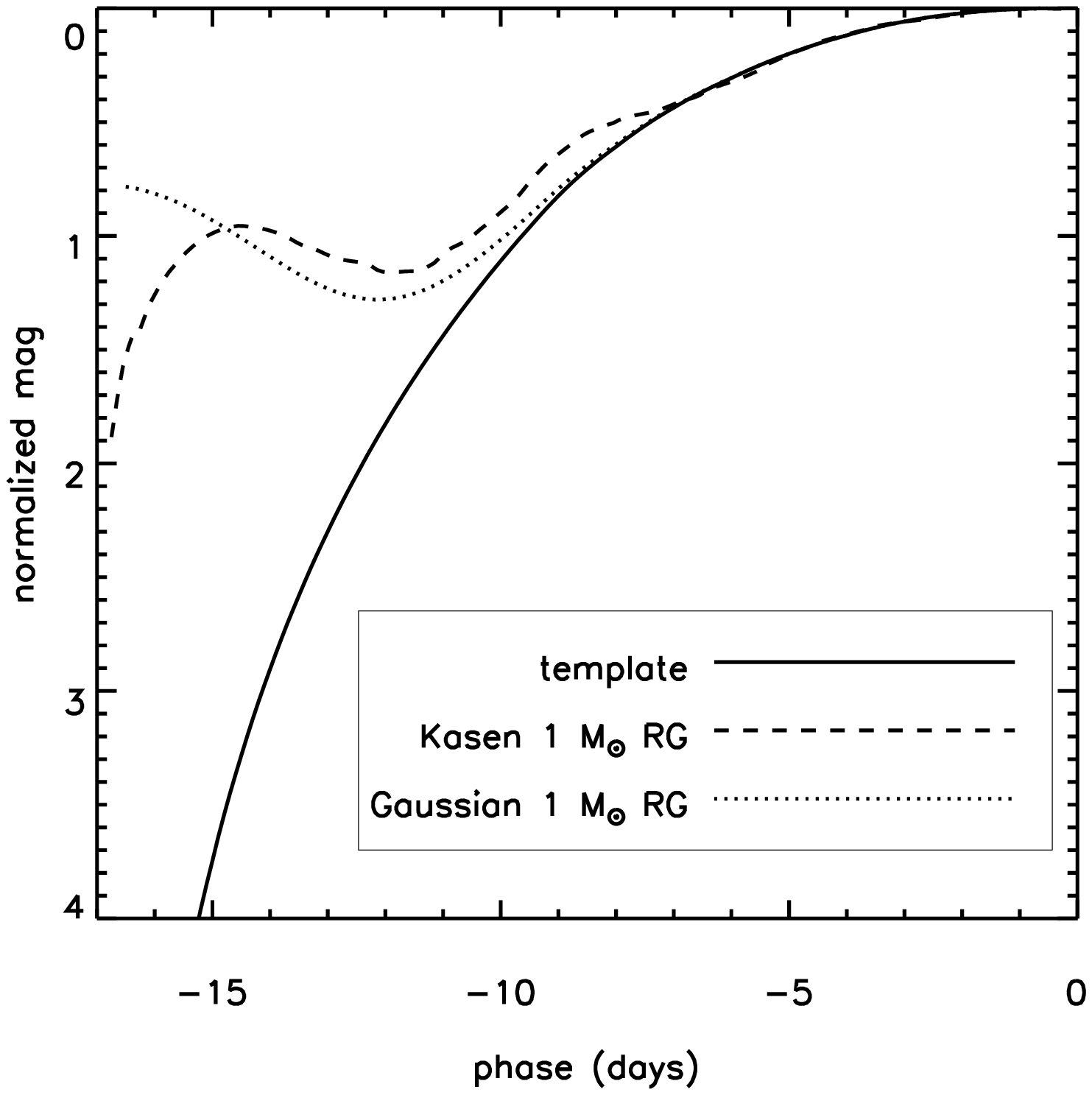}
\end{tabular}
\caption{Model light curves from \citet{kas10} added to the $B$ band MLCS2k2 template, as well as our Gaussian shock model added to the same template. The Gaussian model matches well, and is much easier to produce, making it ideal for our analysis which requires a few million shocks to be produced. The size of the companions are; {\it Top left:} 2 $M_{\odot}$ main sequence, {\it top right:} 6 $M_{\odot}$ main sequence, {\it bottom left:} 1 $M_{\odot}$ red giant. \label{magshock}}
\end{figure}

In Figure \ref{magshock}, we display our simulated shocks in magnitudes compared to the light curves from
Figure~3 of \citet{kas10}. While not a perfect replica of the functional form, our simulations recreate
the overall shape of the theoretical shocks quite well and allow us to vary peak flux and decay time
with a minimum of parameters.

\section{Analysis}

Our analysis in brief proceeds as follows. First, we describe the Brown-Forsythe test of equal variances, then we apply the Brown-Forsythe test to SNANA simulated light curves. In this step, we split our sample of $695$ light curves into two equal groups; one set is shocked at a particular shock width, shock fraction, and shock amplitude, while the other set remains unshocked. For a given set of width, fraction, and amplitude parameters, this test is performed on 1000 different random selections of the two groups, thereby constructing the probability distribution of equal variances for those parameters. This probability distribution is then tested against the distribution where no light curves are shocked at all (in either group) using a student's t-test for equal means. We use this information to construct a probability contour plot for the detectability of shocks based on the cadence and noise properties of the SDSS-II. 

\subsection{The Brown-Forsythe Test}
To compare our simulated shocks to the SDSS-II data we use the Brown-Forsythe test \citep{bf74}. The Brown-Forsythe test is an ANalysis Of VAriance (ANOVA) which compares the absolute deviations from the median between multiple data sets. The computed F statistic is given by
$$ F = \frac{(N-g) \sum_{j=1}^g n_j (z_{.j}-z_{..})^2}{(g-1)\sum_{j=1}^g \sum_{i=1}^{n_j}(z_{ij}-z_{.j})^2} ,$$
where $z_{ij}=|y_{ij}-\tilde{y_j}|/\sigma_{ij}^2$, $\tilde{y_j}$ is the median of the $j^{\rm th}$ group, $g$ is the number of groups, 
$N$ is the total number of data points, $n_j$ is the number of data points in each group, and a `.' in the
subscript indicates the mean over the entire range of subscripts. For example, $z_{.j}$ is the mean of
all deviations in group $j$, and $z_{..}$ is the mean of all deviations in all data sets. By transforming
the response variable of the normal ANOVA test, the Brown-Forsythe test produces an F-statistic which provides
a probability that the variance between the data groups is equal. 

To make sure that the error in each data point is taken into account, the response variable was weighted by $\sigma_{ij}^{-2}$. We tested the weighting with values of $\sigma_{ij}^{-2}$, $\sigma_{ij}^{-1}$, as well as no weighting at all. We found that the type I error rate (percentage of false detections) was smallest for weights of $\sigma_{ij}^{-2}$ when using the SNANA simulated light curves, and chose this for our weights on the transformed response variable. Note that this $\sigma_{ij}$ is the error on each data point and not the shock width, $\sigma$, as defined earlier.

We selected the Brown-Forsythe test because it had a very low type I error rate (false detection) when searching for simulated shocks. After using a number of statistical tests based on the mean of the data, as well as the deviation of each point from mean or median, we settled on the Brown-Forsythe test because it was the most effective and trustworthy in detecting our simulated shocks. Most other statistical test resulted in a large number of false detections when testing simulated shocks, increasing our limits on detectability if those tests were utilized in place of the Brown-Forsythe test.

\subsection{Applying the Brown-Forsythe Test to Simulated Data}
We compare two sets of data at a time (i.e. SNANA shocked vs. SNANA unshocked), in a region from explosion to 3~days after explosion. Testing indicated that this small window was necessary to maintain an acceptable signal-to-noise ratio for the simulated shocks. We compute the median of the data in 1-day bins, and perform a linear interpolation between each median point. We then examine the absolute deviation from this ``median line" using the Brown-Forsythe test. This test is performed on the composite light curve consisting of all light curves in the group, not on individual light curves. Selecting bins smaller than 1 day causes a rare problem (but significant when performing a million random selections) where the random selection of light curves results in a small number of points (less than 5) in one of these bins, yielding a poorly constrained median value. Using bins greater than 1 day wide causes the linear interpolation to be a poor representation of the fast-changing early light curve. 

We produced 1000 simulations for each set of parameters (width, amplitude, and shock fraction) in order to construct
the probability distribution for equal variances between a `shocked' and an `unshocked' sample. This probability
distribution is then compared to other probability distributions (with different shock parameters), using IDL\footnote{Interactive Data Language, http://www.ittvis.com/idl/idl7.asp} to compute the student's t-test probability of equal means. 

\begin{figure}[h!]
\includegraphics*[scale=0.6]{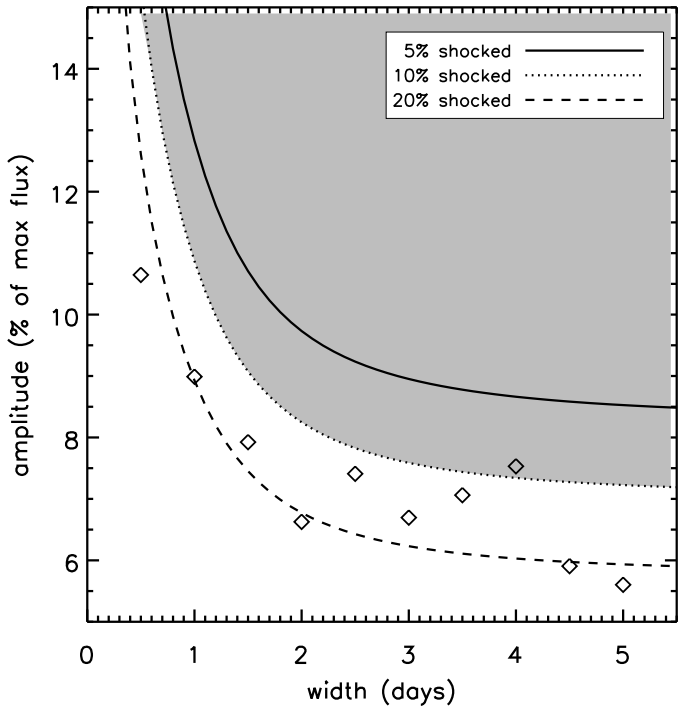}
\caption{Probability contours for 5\%, 10\%, and 20\% shock fractions using the SNANA simulated light curves. The method for 
producing these is described in section 3. The lines represent the 5\% level of probability in parameter space; below the lines, 
shocks are considered undetectable. The shaded region indicates the section of parameter space that we can rule out 
at the 95\% confidence level if the shocks are visible in 10\% of SNe Ia. For over 1200 different combinations of shock fraction, width, and amplitude, 1000 random selections were performed in order to split the 695 SNANA simulated light curves into two groups. One group had a random fraction of the SNe shocked with a fixed width and amplitude. This distribution of 1000 probabilities was then compared to a 1000 point distribution where {\it neither} group was shocked. For a fixed shock fraction and shock width, the probability as a function of amplitude (in log space) is linear until a certain ``cutoff value."  The diamonds are the amplitude and width values (for a fixed 20\% shock fraction) for which the probability drops below 5\%, indicating that even though we have added simulated shocks, our method is unable to detect them due to noise and cadence.  The lines are the functional form given in section 3. The 5\% and 10\% points are not shown; only the fits are shown here.   \label{probcon}}
\end{figure}

For shock fractions of 5\%, 10\% and 20\%, 1000 simulations were performed at a range of shock widths and amplitudes.
For each individual simulation, 695 SNANA simulated curves were randomly split into two approximately equal number groups, 
and a fraction of one set (the ``shock fraction") was artificially shocked. The two sets were then compared using the Brown-Forsythe test described 
above. The widths range from 0 to 5 days in 0.5 day intervals, and the amplitudes range from 0\% to 20\% in 0.5\% intervals. We 
then use the student's t-test to compare each set of 1000 simulated shocks to the case where both SN sets are un-shocked. In log 
space, the probability as a function of amplitude (for fixed width and shock fraction), is relatively constant until a characteristic ``cutoff point", 
where it decreases linearly. By fitting a line to this decreasing portion, we determine the point in amplitude space where the 
probability of equal means (according to the t-test) drops below 5\%. The resulting probability contour plot (Figure  \ref{probcon}) required over a million simulations to be performed in order to constrain the cutoff values.

Figure \ref{probcon} displays the shock detection limits as
a function of shock amplitude and width. For the nominal shock fraction of 10\%, we expect to detect
shocks in the SDSS-II data for emission as faint as $M_B< -16.9$ (typical SN peak of $-19.3$ mag) and lasting a day
or longer.  We approximate a functional form to the shock detection threshold as a function of
shock width and amplitude and fit it to the simulations; 
the functional form of the detection boundary is roughly 
$$ S(\sigma) =A \left(\int_{0}^{3}P(\sigma,t)dt\right)^{-1/2},$$
where $S(\sigma)$ is the shock amplitude as a function of the shock width, $\sigma$. $A$ is a constant
that scales the distribution and
is determined by fitting the function to the simulation results. $P(\sigma,t)$ is the half-Gaussian distribution and $t$
is the time in days since explosion. As expected, the amplitude required for shock detection
increases rapidly as the shock width goes to zero. Beyond a width of about 2~days, the function remains fairly constant
since the detection window is only 3~days. For a shock fraction of 10\%\ and shocks lasting a few days, the shock amplitude must
be above 7\%\ to be detected. Should the shock fraction be closer to 5\%, possibly due to an equal mixture of
DD and SD progenitor systems, then the shocks must peak above 9\%\ to be detectable.

As indicated by Figure \ref{probcon}, this analysis method is able to detect fairly weak shocks,
as we are sensitive to shock emission 2.5~mag fainter than the peak supernova brightness.
The area below the curves represents the region of parameter space where
shocks would be undetectable. To clarify, our recipe for determining the detection limits shown in Figure 3 can be summarized as:
\begin{itemize}
\item Fit all light curves with the 2-stretch fitter, convert them to rest-frame, and 2-stretch correct them to match the template
\item Randomly split the full sample in half, and artificially shock a fraction of one of the two groups
\item Create a composite light curve for each of the two groups, and calculate the deviation from median for every observation in each group
\item Calculate the Brown-Forsythe statistic between the two groups, which uses these median deviations as the variable
\item Perform 1000 random selections  of the groups and shocked SNe, thereby constructing the probability distribution of equal variances for the shock parameters used
\item Compare this distribution of 1000 probabilities to any other distribution (i.e. one where both groups are unshocked) using the student's t-test
\item For a fixed shock fraction and shock width, calculate the ``cutoff value" where the Brown-Forsythe probability drops below 5\%, indicating a detection of shocks. These are the points shown in Figure 3, with the functional form of the boundary fit and then displayed as a line
\end{itemize}

\section{Discussion}
Comparing the SDSS-II data to the un-shocked SNANA simulations gives a Brown-Forsythe
probability of equal variances of 27.4\%. Therefore, we cannot reject the null hypothesis
that the variances are equal. We conclude that the SDSS-II data do not contain shocks above the
levels shown by Figure \ref{probcon}. Comparing the SDSS-II light curves against a random selection of half of
the SNANA un-shocked curves 1000 times produces 19 probability values of less than 5\%. This
is consistent with the statistical type~I error rate (percentage of false positives) determined by comparing two un-shocked
SNANA groups 1000 times, confirming that we do not detect shocks above levels constrained by the simulations.

Another test that was performed was a shock fraction of only $1\%$, with an amplitude of 0.5 and a width of 3.5 days. This corresponds to a scenario where $1\%$ of SNe Ia are single degenerates with a red giant companion. We randomly selected SNe to shock and compared them to unshocked SNe 1000 times using the Brown-Forsythe test, and compared this distribution to 1000 random selections with no shocked light curves (the same method described in section 3). Using the Student's t-test we find a probability less than 2\% that the groups have the same mean probability value, indicating that our method can detect such a small fraction if they are red giant companions. We caution, however, that our analysis cannot rule out red giants in this scenario with only 108 light curves, as we would typically expect to find shock emission from one red giant companion in this group.

Lastly, we performed a test using SNANA shocked curves against the SDSS-II data. If the SDSS-II data contained shock emission, we would expect the student's t-test probability of equal means to decrease as a function of shock amplitude in the SNANA sample. We find that the SNANA shocked sample (shock fraction $10\%$ and width 3.5 days) and the SDSS-II data are indistinguishable statistically until shock amplitudes of 7\%-10\% of peak flux, fully consistent with the scenario involving SNANA shocked curves and SNANA unshocked curves, and reaffirming our finding that shocks are not detected above this level in the SDSS-II data.

Using the results of this analysis, we can constrain properties of companion
stars to SNe~Ia. Employing the equations from \citet{kas10} for $L_{\rm c,iso}$ and $T_{\rm eff}$
(the luminosity of the shock and the effective temperature of the shock) and assuming a
blackbody spectrum, we calculate the theoretical flux density of emission at  4440 \AA ($B-$band).
The flux shortly after explosion (0.1 days) is converted to the corresponding flux density at 10~pc,
and compared to the flux density of a normal SN Ia at 10~pc ($B = -19.3$ mag). This is the
amplitude of the shock as a fraction of SN peak brightness, and it results in an approximate relationship
between $a_{13}$ (stellar separation normalized to $10^{13}$ cm) and the amplitude of the shock, $S$, given by 
$$ S = 0.262\ a_{13}^{0.92}.$$
For a red giant, $a_{13} \approx 2.0$, resulting in a shock amplitude of 0.49, which is easily excluded by our analysis.
The $a_{13}$ of a 6~$M_{\odot}$ main sequence companion star is around 0.2, corresponding to
a shock amplitude of 6\% of peak brightness (these $a_{13}$ values are rough associations; these values assume companions are at the Roche lobe limit). Based on our analysis of simulations and the SDSS-II
data, we would detect shocks in the data if the companions to Type Ia SNe were on average greater
than 6~$M_{\odot}$. This assumes that all SNe Ia occur in the SD channel. If, for example, half
of the observed SNe Ia occur through the DD channel, then the shock fraction decreases to 5\%\ and
our limit moves to larger shock amplitudes (more massive MS stars).

One possible uncertainty in the models is the density distribution in the outer layers of the ejecta. Its effect on the shock light curve can be seen in equation 21 of \citet{kas10}; essentially a shallower density gradient would cause a narrower shock (i.e. the width of the shock would be less). \citet{kas10}  used a fairly steep density profile, however and it corresponds to a width of approximately 3.5 days in our Gaussian shock model. Shallower density profiles could reduce the width and make observing shocks much more difficult (see Figure 3).

\section{Conclusions}
In this study, we simulate the shock interaction of a SN~Ia expected in the single degenerate channel. 
The theory was developed by \citet{kas10}, and is similar to shock breakout in core collapse supernovae. This shock appears as an increase in flux shortly after explosion,
decaying over the course of a couple of days. To simulate this phenomenon, we add a shock light curve that estimates the \citet{kas10} shocked 
light curve (with the ability to vary the amplitude and width) to SNANA simulations at the time of explosion. This approach allows for the setting of constraints on the 
detectability of shocks in the SDSS-II data. Assuming 10\%\ of the shocks are visible to a fixed observer,
then shocks with amplitudes less than 7\% to 9\% would not be detected in the SDSS-II data due
to photometric noise and cadence of the survey.  

Using the Brown-Forsythe test of equal variances, we do not observe shocks in the
SDSS data. This is calculated both directly, with a probability of 27.4\% for equal variance, 
and indirectly, as 1000 random selections comparing the SDSS-II data to the SNANA data
produced only 19 detections, consistent with the $\approx$1.5\% type~I error rate determined
by self-testing the SNANA data. 

From our simulations and the resulting limit on the shock amplitude, the typical companion stars to Type Ia SNe (assuming the 
single degenerate channel make up a considerable fraction of SNe Ia) can be no more than $\sim 6~M_{\odot}$, and red giants are 
ruled out as common companions to SNe~Ia. This is not a particularly tight constraint for the evolution or nature of the progenitor system, considering the white dwarf itself must have been more massive than the companion but less massive than $\sim 7-8~M_{\odot}$. It does however strongly disfavor red giants as usual companions to SNe Ia.

Our limits on shock emission could be improved by designing a survey with a faster cadence or a search sensitive to the ultraviolet or X-rays. An increase in the number of light curves with early data would also improve the significance of this type of testing. It may very well be possible to constrain the progenitors of Type Ia supernovae using shock emission if future surveys are able to meet these criteria. 

\acknowledgements
Funding for the SDSS and SDSS-II has been provided by the Alfred P. Sloan Foundation, the Participating Institutions, the National Science Foundation, the U.S. Department of Energy, the National Aeronautics and Space Administration, the Japanese Monbukagakusho, the Max Planck Society, and the Higher Education Funding Council for England. The SDSS Web Site is http://www.sdss.org/.

The SDSS is managed by the Astrophysical Research Consortium for the Participating Institutions. The Participating Institutions are the American Museum of Natural History, Astrophysical Institute Potsdam, University of Basel, University of Cambridge, Case Western Reserve University, University of Chicago, Drexel University, Fermilab, the Institute for Advanced Study, the Japan Participation Group, Johns Hopkins University, the Joint Institute for Nuclear Astrophysics, the Kavli Institute for Particle Astrophysics and Cosmology, the Korean Scientist Group, the Chinese Academy of Sciences (LAMOST), Los Alamos National Laboratory, the Max-Planck-Institute for Astronomy (MPIA), the Max-Planck-Institute for Astrophysics (MPA), New Mexico State University, Ohio State University, University of Pittsburgh, University of Portsmouth, Princeton University, the United States Naval Observatory, and the University of Washington.

This work is based in part on observations made at the 
following telescopes.
The Hobby-Eberly Telescope (HET) is a joint project of the University of Texas
at Austin,
the Pennsylvania State University,  Stanford University,
Ludwig-Maximillians-Universit\"at M\"unchen, and Georg-August-Universit\"at
G\"ottingen.  The HET is named in honor of its principal benefactors,
William P. Hobby and Robert E. Eberly.  The Marcario Low-Resolution
Spectrograph is named for Mike Marcario of High Lonesome Optics, who
fabricated several optical elements 
for the instrument but died before its completion;
it is a joint project of the Hobby-Eberly Telescope partnership and the
Instituto de Astronom\'{\i}a de la Universidad Nacional Aut\'onoma de M\'exico.
The Apache 
Point Observatory 3.5 m telescope is owned and operated by 
the Astrophysical Research Consortium. We thank the observatory 
director, Suzanne Hawley, and site manager, Bruce Gillespie, for 
their support of this project.
The Subaru Telescope is operated by the National 
Astronomical Observatory of Japan. The William Herschel 
Telescope is operated by the 
Isaac Newton Group, and the Nordic Optical Telescope is 
operated jointly by Denmark, Finland, Iceland, Norway, 
and Sweden, both on the island of La Palma
in the Spanish Observatorio del Roque 
de los Muchachos of the Instituto de Astrofisica de 
Canarias. Observations at the ESO New Technology Telescope at La Silla
Observatory were made under program IDs 77.A-0437, 78.A-0325, and 
79.A-0715.
Kitt Peak National Observatory, National Optical 
Astronomy Observatory, is operated by the Association of 
Universities for Research in Astronomy, Inc. (AURA) under 
cooperative agreement with the National Science Foundation. 
The WIYN Observatory is a joint facility of the University of 
Wisconsin-Madison, Indiana University, Yale University, and 
the National Optical Astronomy Observatories.
The W. M. Keck Observatory is operated as a scientific partnership 
among the California Institute of Technology, the University of 
California, and the National Aeronautics and Space Administration. The 
Observatory was made possible by the generous financial support of the 
W. M. Keck Foundation. 
The South African Large Telescope of the South African Astronomical 
Observatory is operated by a partnership between the National 
Research Foundation of South Africa, Nicolaus Copernicus Astronomical 
Center of the Polish Academy of Sciences, the Hobby-Eberly Telescope 
Board, Rutgers University, Georg-August-Universit\"at G\"ottingen, 
University of Wisconsin-Madison, University of Canterbury, University 
of North Carolina-Chapel Hill, Dartmouth College, Carnegie Mellon 
University, and the United Kingdom SALT consortium. The Italian Telescopio Nazionale Galileo is 
operated on the island of La Palma by the Fundacion Galileo Galilei of the Instituto Nazionale
di Astrofisica at the Spanish Observatorio del Roque de los Muchachos
of the Instituto de Astrofisica de Canarias. 

Support for this research at Rutgers University was provided by DOE
grant DE-FG02-08ER41562 and NSF CAREER award 0847157 to SWJ.

\end{document}